\documentclass [11pt] {article}

\usepackage[dvips]{graphicx,color}

\usepackage{verbatim}

\newcommand{\half}{{\textstyle{1\over2}}}
\newcommand{\dt}{\Delta t}
\newcommand{\inv}[1]{\frac{1}{#1}}
\newcommand{\E}[1]{e^{{#1}}}
\newcommand{\der}[3]{\frac{\partial^{#1} #2}{\partial{#3}}}
\newcommand{\bra}{\langle} 
\newcommand{\ket}{\rangle} 
\newcommand{\bfx}{{\bf x}}
\newcommand{\bfX}{{\bf X}}
\newcommand{\bfi}{{\bf i}}
\newcommand{\bfk}{{\bf k}}
\newcommand{\bff}{{\bf f}}
\newcommand{\bfn}{{\bf n}}
\newcommand{\bfp}{{\bf p}}
\newcommand{\bfone}{{\bf 1}}
\newcommand{\bfeta}{{\mbox{\boldmath $\eta$}}}
\newcommand{\bfOmega}{{\mbox{\boldmath $\Omega$}}}
\newcommand{\mod}{{\mbox{mod}}}

\makeatletter
\@addtoreset{equation}{section}
\makeatother
\renewcommand{\theequation}{\thesection.\arabic{equation}}
\def\appendix{\par        
 \setcounter{section}{0}      
 \setcounter{subsection}{0}
 \renewcommand{\theequation}{\Alph{section}.\arabic{equation}}
 \renewcommand{\thesection}{Appendix
 \Alph{section}\setcounter{equation}{0}}
}
 
\catcode`\@=11
 
\def\applabel#1{\@bsphack
 \protected@write\@auxout{}%
   {\string\newlabel{#1}{{\Alph{section}}{\thepage}}}%
 \@esphack}

\begin{document} 

\title{
  \begin{flushleft}
    {\footnotesize BU-CCS-990401}\\
    {\footnotesize LA-UR 00-118}\\[0.5cm]
  \end{flushleft}
  {\bf Fourier Acceleration of Langevin Molecular Dynamics}
  }

\author{
  Francis J. Alexander\\
  {\footnotesize CIC-3, MS-B256, Los Alamos National Laboratory, University of California}\\
  {\footnotesize Los Alamos, New Mexico 87545}\\
  {\footnotesize{\tt fja@lanl.gov}}\\[0.3cm]
  Bruce M. Boghosian and Richard C. Brower\\
  {\footnotesize Center for Computational Science and Department of Physics, Boston University,}\\
  {\footnotesize 3 Cummington Street, Boston, Massachusetts 02215}\\
  {\footnotesize{\tt bruceb@bu.edu} and {\tt brower@bucrf20.bu.edu}}\\[0.3cm]
  S. Roy Kimura\\
  {\footnotesize Department of Biomedical Engineering, Boston University,}\\
  {\footnotesize Boston, Massachusetts 02215}\\
  {\footnotesize{\tt srk@engc.bu.edu}}
  }

\maketitle

\begin{abstract}
  Fourier acceleration has been successfully applied to the simulation
  of lattice field theories for more than a decade.  In this paper, we
  extend the method to the dynamics of discrete particles moving in
  continuum.  Although our method is based on a mapping of the
  particles' dynamics to a regular grid so that discrete Fourier
  transforms may be taken, it should be emphasized that the introduction
  of the grid is a purely algorithmic device and that no smoothing,
  coarse-graining or mean-field approximations are made.  The method
  thus can be applied to the equations of motion of molecular dynamics
  (MD), or its Langevin or Brownian variants.  For example, in Langevin
  MD simulations our acceleration technique permits a straightforward
  spectral decomposition of forces so that the long-wavelength modes are
  integrated with a longer time step, thereby reducing the time required
  to reach equilibrium or to decorrelate the system in equilibrium.
  Speedup factors of up to 30 are observed relative to pure
  (unaccelerated) Langevin MD.  As with acceleration of critical lattice
  models, even further gains relative to the unaccelerated method are
  expected for larger systems.  Preliminary results for
  Fourier-accelerated molecular dynamics are presented in order to
  illustrate the basic concepts.  Possible extensions of the method and
  further lines of research are discussed.
\end{abstract}

\newpage

\section{Introduction}

Molecular dynamics (MD) simulations play an important role in our
fundamental understanding of the kinetics of molecular systems and
provide a powerful tool for modeling a wide variety of materials
including biomolecules.  Although MD simulations have benefited
tremendously from advances in high-performance computing, they suffer
from the limitation arising from the numerical stiffness inherent in
Newton's equations.  The result is that MD studies are generally
restricted to short intervals of real time, from nanoseconds up to a few
microseconds, even with heroic computational efforts.  To overcome this
difficulty, there is a growing effort to develop accelerated MD
algorithms.  (See, for example,
\cite{rabitz,berne,schlick,voter,bornemann,jensen}. )

In contrast to molecular (or other discrete particle) systems with a
Lagrangian data representation, there is a considerable variety of
acceleration algorithms available for continuum field theories
approximated on a regular grid or lattice.  For example, grid-based
simulations have made substantial progress with the advent of cluster
Monte Carlo methods~\cite{swang,wolff}, Fourier acceleration
\cite{batrouni}, and multi-grid iterative solvers~\cite{wesseling}.
Because the bulk properties of large aggregates of molecules can often
be described by continuum mechanics, it is intuitively appealing that a
corresponding method should apply in the molecular (or particulate)
framework.  Indeed, making this connection between the molecular and
continuum scales is a central goal for multi-scale modeling projects.
In this paper we show how one such continuum tool, namely Fourier
acceleration, can be applied to Langevin MD without introducing any
coarse-graining or mean-field approximations.  The basic ingredient of
the method is an exact mapping of the original particulate system onto a
regular lattice of displacement fields.  Although this mapping may prove
useful in a broader context, we restrict our attention to Fourier
acceleration of the Langevin equations for MD.

The idea of introducing a regular grid into MD is not new; grid-based
recursive multipole expansions~\cite{multipole}, for example, have been
used for more than a decade to rapidly compute Coulomb interactions.
More recently, hybrid atomistic-continuum techniques, such as the
quasi-continuum method~\cite{quasicontinuum}, use finite-element
techniques to bridge microscopic and macroscopic length scales.  Most
applications of spectral methods to molecular systems, however, have
been confined to the analysis of data (for example, structure and
response functions).

By contrast, our procedure uses spectral analysis to modify and
accelerate the dynamical evolution of the molecular system.  Unique to
this approach is the mapping of the actual position coordinates to a
grid, and the ability to invert the mapping to displace the original
off-lattice molecular coordinates.  The introduction of the grid is a
purely algorithmic device and is not tantamount to a coarse-graining or
mean-field approximation of any kind; that is, the accelerated dynamics
are still those of discrete particles.  The result is a new,
accelerated, stochastic dynamics that is significantly faster than
standard Langevin MD, but still exactly preserves the equilibrium
distribution.  The fundamental tradeoff associated with this approach is
that of speed versus faithfulness to the essential kinetics.  Both of
these desiderata are clearly specific to the system being studied and
the phenomena that the model should faithfully represent.

The organization of this paper is as follows.  Section~\ref{sec:maps}
describes our procedure for mapping the particulate system to a regular
lattice.  This mapping is a prerequisite to the application of a
Fourier-mode decomposition.  In Section~\ref{sec:fald} we outline the
Fourier-accelerated Langevin dynamics on the grid.  We demonstrate the
method in Section~\ref{sec:phi} by applying it to a $\phi^4$ model at
its critical point.  We then describe how to apply Fourier acceleration
(FA) in conjunction with the lattice mapping in Section~\ref{sec:famd}.
As an example, in Section~\ref{sec:tdlj}, we apply the method to the
Langevin dynamics of a simple Lennard-Jones fluid.  Extensions of the
Fourier-accelerated molecular dynamics (FAMD) method and additional
applications are discussed in Section~\ref{sec:disc}.

\section{Particle-to-Grid Mapping Procedure}
\label{sec:maps}

There are many ways by which a molecular system can be transformed from
(off-lattice) particle coordinates to a fiducial grid.  Each has its
advantages and disadvantages, depending upon the aim of the
transformation.  In this section we discuss one method that has proven
to be particularly useful.

In one dimension, the simplest mapping procedure procedure is to sort
the particles by their position coordinate.  Each particle $i$ is given
a permuted label $n(i)$ so that $n(i) < n(j)$ if $x_i < x_j$.  Whereas
the $i$ and $j$ indices are arbitrary labels, devoid of physical
significance, the permuted labels are based on the sequence of particle
positions and hence may be thought of as lying on a grid with some
physical meaning.  Because $n(i)$ is a permutation, it has inverse
function $i(n)$ such that $n(i(\cdot))=\cdot$.  We now transform to new
coordinates by the prescription $X_n = x_{i(n)}$.  This mapping makes it
possible to directly Fourier transform the new position coordinates,
\begin{equation}
\tilde{X}_k = \inv{L} \sum_n X_n \E{i k n} . 
\end{equation}
Note that this new spatial representation contains precisely the same
amount of information as the original data.  Also note that because this
mapping is merely a geometrically motivated relabeling, all attributes,
such as mass $m_i$, are automatically transfered to the new
representation.

The multidimensional generalization of this method is not so
straightforward.  The problem of sorting the particles in more than one
dimension is not well defined.  There is, however, a very good and
efficient approximation used by numerical analysts for load-balancing
graphs on multiprocessor architectures, known as {\it Recursive
  Coordinate Bisection} (RCB)~\cite{simon}.  To see how this algorithm
works, consider a two-dimensional square domain containing $N = L^2$
particles, where $L$ is a power of 2.  We first introduce a two-index
label $\bfi=(i_x,i_y)$ for the particles, where
\begin{eqnarray}
i_x(i) &=& i\;\mod\; L\\
i_y(i) &=& (i - i_x)/L,\\
\noalign{\noindent so that}\nonumber\\
i(\bfi) &=& i_x + (i_y-1)L.
\end{eqnarray}
Thus the transformation from one-index labels $i$ to two-index labels
$\bfi$ is a bijection.  We can label the particles' coordinates as
$\bfx_i=(x_i,y_i)$, or equivalently as
$\bfx_\bfi=(x_\bfi,y_\bfi)=(x_{i(\bfi)},y_{i(\bfi)})$.  as with the $i$
labels for the one-dimensional case, these labels (both $i$ and $\bfi$)
are assigned arbitrarily and devoid of physical content.

Whereas it is difficult to see how to order the one-index labels $i$,
the RCB method provides a straightforward prescription for permuting the
two-index labels $\bfi$ into a new set of two-index labels $\bfn(\bfi)$
that are based on the particles' positions.  Again, this function is a
permutation, so it has an inverse $\bfi(\bfn)$, such that
$\bfn(\bfi(\cdot))=\cdot$.  The $\bfn$'s may reasonably be taken to lie
on a regular two-dimensional grid, and hence provide a set of
independent variables with respect to which the new coordinates
$\bfX_\bfn\equiv\bfx_{\bfi(\bfn)}=\bfx_{i(\bfi(\bfn))}$ can be Fourier
transformed, just as in the one-dimensional example.

To accomplish this, the physical domain is first divided into left- and
right-hand portions with equal numbers ($N/2$) of particles, by sorting
the particles on their $x$ coordinates.  The half with smaller $x$
coordinates will have $1 \leq n_x \leq L/2$ and the half with larger $x$
coordinates will have $L/2 < n_x \leq L$.  In binary notation, this
labeling sets the most significant bit of the $n_x$ index to zero/one
for the left/right halves.  Next, we sort each set of $N/2$ particles by
their $y$ coordinates to obtain 4 sets of $N/4$ particles, likewise
setting the most significant bit in the $n_y$.  This procedure is then
applied recursively to each of the 4 boxes with $N/4$ particles,
maintaining the alternation between the $x$ and $y$ axes.  For systems
of relatively uniform density, the resulting fiducial grid leads to a
remarkably regular and local particle-labelling scheme.  RCB is an order
$N\log N$ algorithm with many obvious similarities to fast Fourier
transforms.

\begin{figure}
  \center{
    \mbox{
      \includegraphics[bbllx=72,bblly=164,bburx=540,bbury=630,width=5.0truein]{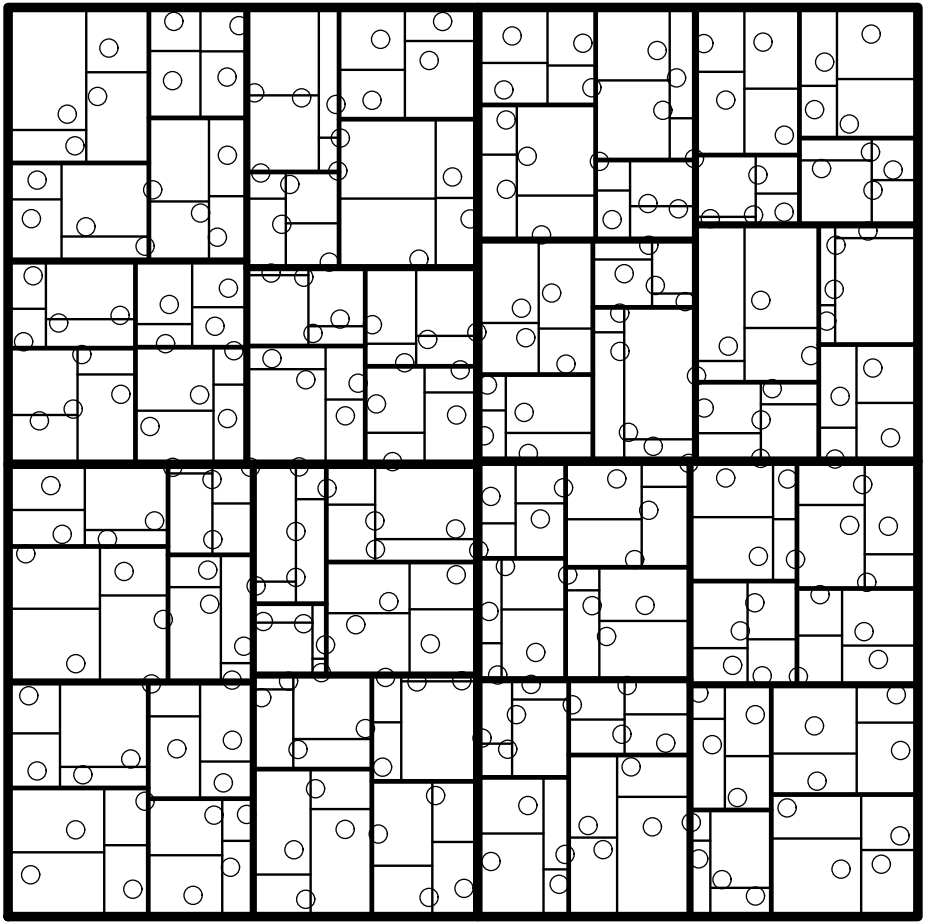}
      }}
  \caption{Grid Mapping for a two-dimensional Lennard Jones fluid by
    Recursive Coordinate Bisection.}
  \label{fig:rcb}
\end{figure}

\section{Fourier Accelerated Langevin Dynamics}
\label{sec:fald}

To demonstrate how Fourier Acceleration works, we consider in detail a
simple (discrete-time) Langevin dynamics.  The Langevin equation of
motion for a system of $N$ particles is
\begin{equation} 
\bfx_i(t+\Delta t) = \bfx_i(t) +
\frac{\bff_i(t)}{2m_i}\left(\dt\right)^2 +
\bfp_i(t)\dt,
\end{equation}
where the $N$ momenta are Gaussian random variables
$\bra\bfp_i(t)\bfp_j(t')\ket = \half k_BT m_i \delta_{i,j}
\delta_{t,t'}\bfone$.  It is well known that this dynamics (in the limit
of vanishing time step) samples the canonical-ensemble Boltzmann-Gibbs
equilibrium distribution function,
\begin{equation} 
P(\bfx_i,\bfp_i) =
\inv{Z}
\exp
\left[
  -\beta
  \left(
    \frac{\beta p_i \cdot p_i}{2 m_i} +
    V(\bfx_i)
  \right)
\right],
\end{equation}
where $\beta\equiv 1/k_BT$ and the force is $f_i = -\partial V/ \partial
\bfx_i$.

For the moment, we set this result aside and consider lattice field
problems for which Batrouni et al.~\cite{batrouni} have shown how to
accelerate the approach to equilibrium in Fourier space.  For example,
consider fields $\phi_\bfx(t)$ on a uniform grid with sites $\bfx$, obeying
the equilibrium distribution,
\begin{equation}
\label{eq:equil}
 P(\phi_\bfx) = \inv{Z} \E{ - S(\phi_\bfx)} \; 
\end{equation}
with action $S$.  This distribution is a fixed point of the discrete
Langevin dynamics (as $\dt \to 0$),
\begin{equation}
\phi_\bfx(t+\dt) =
\phi_\bfx(t) - K \der{}{S(\phi_\bfx)}{\phi_\bfx}\left(\dt\right)^2 +
\eta_\bfx(t)\dt \; ,
\label{eq:dyn}
\end{equation} 
where $\eta_\bfx(t)$ are Gaussian random fields.  This Markov process,
however, is not the only one which drives the system to the equilibrium
in Eq.~(\ref{eq:equil}).  Indeed, the local dynamics of
Eq.~(\ref{eq:dyn}) generally exhibits long autocorrelation times near
critical points.  Batrouni et al.~\cite{batrouni} have shown how to
accelerate such grid-based Langevin equations using a Fourier
decomposition of the dynamics.  The Fourier Acceleration (FA) method
depends on the simple observation that {\em any} mobility (or inverse
mass) matrix may be introduced by the substitutions, $ K \rightarrow
K_{\bfx,\bfx'}$ and $ \bra \eta_\bfx(t) \eta_\bfx'(t) \ket =
K_{\bfx,\bfx'} \delta(t',t)\bfone$, without upsetting the equilibrium
distribution of the fields.  One such choice is a matrix $K$ which is
diagonal in Fourier space.  This choice leads to acceleration if the
time steps for the slow (low wavenumber) modes are amplified.
\begin{equation} 
\phi_\bfx(t+\Delta t) =
\phi_\bfx(t) +
{\cal F}^{-1}
\left[
  -\tilde K(\bfk) {\cal F}
  \left[
    \der{}{S}{\phi_\bfx}
  \right]\left(\dt\right)^2 +
  \sqrt{\tilde K(\bfk)} \;\eta_\bfk\;\dt
\right],
\end{equation}
where ${\cal F}$ represents a Fourier transform.  A simple substitution
of the field $\phi_\bfx$ with the position of a particle $\bfx_\bfi$
would allow us to use Fourier methods.  The mappings of
Sec.~\ref{sec:maps} provide that substitution.
 
\section{$\phi^4$ Model}
\label{sec:phi}

To illustrate the above procedure we apply the FA technique to the
$\phi^4$ model at the critical point in two dimensions.  It has been
shown by Batrouni {\em et al.}~\cite{batrouni} that critical slowing
down is completely eliminated by FA in a purely Gaussian model.  Of
course, in that case, the modes completely decouple in momentum space
and each can be integrated independently.  For a nonlinear model with
mode-coupling, it is not guaranteed that FA will work at all.  It is
also not clear {\em a priori} what the optimal choice of the mass matrix
should be that will most rapidly drive the system to equilibrium or
decorrelate the system once in equilibrium.

To gain experience in selecting the mass matrix for FAMD, we first
studied a simpler system, namely the $\phi^4$ model in two dimensions at
criticality.  This model provides a qualitative (and in some cases
quantitative) description of a displacive phase transition.  The
investigation of such transitions is one of our long-term goals.
Surprisingly, the FA method applied to this system at its critical point
has not been analyzed.  However, Batrouni and Svetitsky have studied its
application to first-order phase transitions in a $\phi^4$ model and
found a significant speedup of tunneling between
minima~\cite{batrouni_fopt}.

The Hamiltonian is given by,
\begin{eqnarray}
\beta {\cal
H}([\phi])
&=&
\sum_{i=1}^N
\left[ \frac{-\theta}{2} \phi_i^2 +
\frac{\chi}{4} \phi_i^4 +
\frac{1}{2} \sum_{\mu=1}^d \left(
\phi_{i_\mu} - \phi_i \right)^2 \right] \nonumber \\
&=&
\sum_{i=1}^N
\left[\frac{\tilde{\theta}}{2} \phi_i^2 +
\frac{\chi}{4} \phi_i^4 -
\frac{1}{2} \sum_{\mu=1}^{2d} \phi_i \phi_{i_\mu} \right],
\end{eqnarray}
where
\begin{equation}
\tilde{\theta} = 2d - \theta \;\;. 
\end{equation}
This system exhibits critical behavior along a line of critical
parameters, $\chi$ and $\theta$.  We simulated the system at the
critical point, $\chi = 1. 0$, $\theta = 1. 265$, which was numerically
determined previously by Toral and Chakrabarti~\cite{toral}.

We updated this system using the Fourier accelerated Langevin equation
described above, namely,
\begin{equation}
\phi_i(t+\Delta t) = \phi_i(t) + {\cal F}^{-1} \left[
K(\bfk) {\cal F} \left( -\frac{\delta {\cal H}}{\delta \phi}
\right) + \sqrt{k_BT\; K(\bfk)}\;\;\tilde{\eta}(\bfk)
\right],
\end{equation}
where $\tilde{\eta}(\bfk)$ represents the Fourier transformed
Gaussian noise with $\langle \eta \rangle = 0$ and $\langle \eta^2
\rangle = 1$, and $K(\bfk)$ represents the mass matrix which gives us
the desired acceleration.  We chose the mass matrix to be the lattice
propagator of the free theory,
\begin{equation}
K(\bfk) =
\frac{4d+m^2}{4\sum_{\mu=1}^{d} \sin^2\frac{k_\mu}{2} +
m^2} \;\left(\dt\right)^2\;,
\label{eq:freeprop}
\end{equation}
where the parameter $m$ is expected to be order $1/\xi$ or $1/L$ for
finite-size scaling~\cite{batrouni}.  The value of this parameter was
adjusted during trial simulations by setting $m = c/L$ for different
values of the constant $c$.  We report the results for $c=4\sqrt{2}$.
As a check, we repeated the simulations without Fourier acceleration
using the pure Langevin update,
\begin{equation}
\phi_i(t+\Delta t) = \phi_i(t) + \frac{\Delta
t^2}{2}f(\phi) + \sqrt{k_BT}\;\eta,
\end{equation}
where $\eta$ is a zero-mean unit-variance Gaussian random variable and
the force term, $f(\phi)$, is,
\begin{equation}
f(\phi) = - \frac{\delta{\cal H}}{\delta \phi} = - \sum_{i=1}^{N}
\left(
\tilde{\theta}\phi_i - \chi\phi_i^3 - \sum_{\mu=1}^{2d} \phi_{i_\mu}
\right) . 
\end{equation}

\begin{figure}
  \center{
    \mbox{
      \includegraphics[bbllx=60,bblly=198,bburx=550,bbury=616,width=5.0truein]{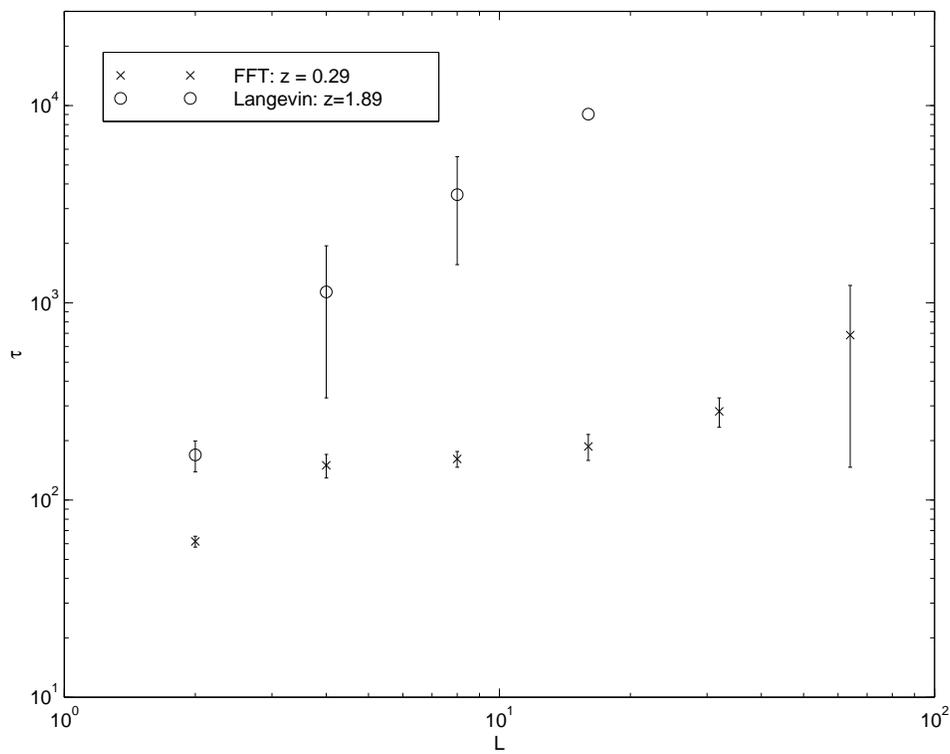}
      }}
\caption{Finite size scaling of the correlation
  time $\tau$ with the linear dimension $L$ for Langevin and FA Langevin
  simulations of $\phi^4$ theory; $\tau$ is in units of $\Delta t$.}
\label{fig:tauL}
\end{figure}

Finite-size scaling simulations were conducted for $L \times L$ system
sizes where $L=2$, 4, 8, 16, 32, 64 using the FA Langevin update and for
$L=2$, 4 , 8, 16 for the pure Langevin case.  A time step of $\Delta
t=0. 05$ was used.  Note that our time step corresponds to
$\sqrt{\epsilon}$ in Ref.~\cite{batrouni} and thus should give very
little discretization error.  We ran each system for time that is
approximately 1000 times longer the correlation time estimated from
trial simulations.  The results are shown in Fig.~\ref{fig:tauL}.  The
normalized correlation functions, $C(t) = \langle E(0)E(t)
\rangle/\langle E(0)E(0) \rangle$, were computed from the time series of
the energy density.  Correlation times $\tau$ for each $L$ were computed
by fitting the region $0. 3 \le C(t) \le 0. 6$ to $\exp(-t/\tau)$.
Error bars were estimated from the standard deviation of the values of
$\tau$ measured from five independent time series per system.  Average
energies and standard deviations of 10 blocked averages were measured
for each system size and update algorithm.  These results are listed in
Table~\ref{tab:mean1}.
 
\begin{table}
\begin{center}
\begin{tabular}{l|rl|rl|rl}
\multicolumn{1}{c|}{System Size} &
\multicolumn{2}{c|}{Heat bath} &
\multicolumn{2}{c|}{Langevin} &
\multicolumn{2}{c}{FA} \\
 & \multicolumn{1}{c}{\scriptsize Value} &
   \multicolumn{1}{c|}{\scriptsize Error} &
   \multicolumn{1}{c}{\scriptsize Value} &
   \multicolumn{1}{c|}{\scriptsize Error} &
   \multicolumn{1}{c}{\scriptsize Value} &
   \multicolumn{1}{c}{\scriptsize Error}\\
\hline
$L=2$ & 0. 481 & $\pm$ 0. 0159 & 0. 496 & $\pm$ 0. 0612 & 0. 473 &
$\pm$ 0. 0432 \\
$L=4$ & 3. 174 & $\pm$ 0. 0411 & 3. 180 & $\pm$
0. 4019 & 3. 291 & $\pm$ 0. 2998 \\
$L=8$ & 14. 54 & $\pm$ 0. 1576 &
14. 65 & $\pm$ 0. 9797 & 14. 87 & $\pm$ 0. 5239 \\
$L=16$ & 61. 16 &
$\pm$ 0. 4790 &  &    & 62. 58 & $\pm$ 1. 6278 \\
$L=32$
& 251. 5 & $\pm$ 1. 6066 &  &    & 260. 6 & $\pm$
3. 5662
\end{tabular}
\end{center}
\caption{Mean energies and errors for each
system size and update method.  Errors are standard deviations
from 10 blocked averages.}
\label{tab:mean1}
\end{table}

In Fig.~\ref{fig:tauL} we compare the autocorrelation times for the pure
Langevin update with those for the Fourier-Langevin case.  There is
clearly an acceleration.  Whether the dynamical exponent $z$, which
describes the growth of autocorrelation times by the scaling relation,
$\tau = L^z$, is actually different for Langevin and Fourier
Acceleration is an interesting and open question, and would require more
extensive computation than we have done to date.  For practical
simulations of systems far from criticality, the value of $z$ is often
not as important as the overall amplitude of the autocorrelation time.
 
\section{Fourier Accelerated Molecular Dynamics}
\label{sec:famd}

To apply these techniques to discrete particles with a Lagrangian data
representation, we must introduce a Fourier transform of the position
coordinates $\bfx_i(t)$.  Clearly we cannot simply transform the
$\bfx_i$ with respect to the particle labels $i = 1,2,\ldots N$.  As
mentioned in Sec.~\ref{sec:maps}, these labels are generally devoid of
physical meaning.  They have no natural relationship to the properties
of the particles or to their spatial and/or temporal configuration.
Hence, the first step is to map the particles onto a uniform spatial
grid.

The mapping scheme discussed in Sec.~\ref{sec:maps} suggests how to
define appropriate grid coordinates.  In the Index Method, the mapping,
$\bfx_i \rightarrow \bfX_\bfn$, is simply a relabling of the
coordinates.  (To be more explicit this notation for a 2D system is
expanded into components: $\bfx_i \equiv (x_i,y_j)$ and $\bfX_\bfn =
(X_{n_x,n_y},Y_{n_x,n_y})$, where $\bfn = (n_x,n_y)$ is a two component
integer vector. ) Consequently the Langevin dynamics is unaffected,
\begin{equation}
\bfX_\bfn(t+\dt) =
\bfX_\bfn(t) +
\frac{\bff_\bfn(t)}{2 m_i}\dt^2 +
\sqrt{k_BT}\; \bfeta_\bfn(t)\dt,
\end{equation}
where we have introduced the normalized independent Gaussian noise with
variance $\bra \bfeta\bfeta \ket = \bfone$.  Because we have established
a two-dimensional grid, we may now try to accelerate the dynamics simply
by going to Fourier space as described above for a generic lattice field
theory.  The fields are now the position vectors of each particle.  As
we will demonstrate numerically in Sec.~\ref{sec:tdlj}, this grid is
indeed useful for a two-dimensional Lennard-Jones fluid.

\section{2D Lennard-Jones Fluid}
\label{sec:tdlj}

Motivated by the successful application of Fourier Acceleration to
decorrelate lattice $\phi^4$ systems, we have tested its ability to
reduce the autocorrelation time of a system of Lennard-Jones atoms in
two dimensions using the Index Method.

The Lennard-Jones interaction potential is given by
\begin{equation}
V_{LJ}(r) = 4 \epsilon \left( \; \frac{\sigma^{12}}{r^{12}} -
\frac{\sigma^{6}}{r^{6}} \right). 
\end{equation}
In our simulations we chose $\epsilon = \sigma = 1$, a potential cutoff
at $2. 5\sigma$, and worked at the liquid-vapor critical point, with
temperature and density parameters $T=T_c=0. 47$, $\rho_c = 0. 35$,
respectively.  Both pure Langevin MD and FAMD were tested for $N= 16$
and 64 particles with periodic boundary conditions.  Each system was
evolved on the order of $10^7$ integration steps with $\Delta t = 0.
005$.  This time step allowed us to accurately determine the critical
thermodynamic quantities.  The acceleration kernel we used in FAMD was
identical in form to the one we applied to the $\phi^4$ model, namely
\begin{equation}
\epsilon(\bfk) =
\frac{4d+\frac{1}{N}}{4\sum_{\mu=1}^{d} \sin^2\left(\frac{k_\mu}{2}\right) +
\frac{1}{N}}\; \left(\Delta t\right)^2 \;,
\end{equation} 
where $N$ is now the number of particles in the system.  This should be
compared to Eq.~(\ref{eq:freeprop}).

We allowed the system to evolve for $10^6$ steps before statistics were
taken.  To compare the effectiveness of the Fourier acceleration, we
examined the autocorrelation of various long-wavelength modes of the
system.  In particular, we looked at the circularly averaged
time-autocorrelation of the cosine-transformed density.  In $D$ spatial
dimensions, we write $d\bfk = k^{D-1}dk\;d\bfOmega$, where $k=|\bfk|$
and $d\bfOmega$ is the direction differential, so the cosine-transformed
density is
\begin{equation}
\rho(\bfk,t) = \sum_i^{N}\cos (\bfk\cdot\bfx_i).
\end{equation}
The autocorrelation that we measure is then
\begin{equation}
A(k,\tau)
\equiv
\frac{\int d\bfOmega\; \rho(\bfk,t+\tau)\rho(\bfk,t)}{\int d\bfOmega},
\end{equation}
where $\bfk=2\pi\bfn/L$, and $L$ is the linear dimension of the system.

\begin{table}[t]
\begin{center}
\begin{tabular}{l|rl|rl}
\multicolumn{1}{c|}{Wavenumber} &
\multicolumn{2}{c|}{FAMD} &
\multicolumn{2}{c}{Langevin} \\
 & \multicolumn{1}{c}{\scriptsize Value} &
   \multicolumn{1}{c|}{\scriptsize Error} & 
   \multicolumn{1}{c}{\scriptsize Value} &
   \multicolumn{1}{c}{\scriptsize Error}\\
\hline
$n=1$ & 12000 & $\pm$ 3000 & 80000 & $\pm$ 12000 \\ 
$n=2$ & 3300 & $\pm$ 500 & 22000 & $\pm$ 4000 \\
$n=4$ & 1100 & $\pm$ 130 & 8600 & $\pm$ 600 \\
\end{tabular}
\end{center}
\caption{Correlation times in (integration time steps) for different
  wavelengths for $N=16$ particles.}
\label{tab:mean2}
\end{table}

\begin{table}[t]
\begin{center}
\begin{tabular}{l|rl|rl}
\multicolumn{1}{c|}{Wavenumber} &
\multicolumn{2}{c|}{FAMD} &
\multicolumn{2}{c}{Langevin} \\
 & \multicolumn{1}{c}{\scriptsize Value} &
   \multicolumn{1}{c|}{\scriptsize Error} &
   \multicolumn{1}{c}{\scriptsize Value} &
   \multicolumn{1}{c}{\scriptsize Error}\\
\hline
$n=1$ & 40000 & $\pm$ 13000 & 1200000 & $\pm$ 130000 \\ 
$n=2$ & 15000 & $\pm$ 2000 & 310000 & $\pm$ 60000 \\
$n=4$ & 4200 & $\pm$ 700 & 40000 & $\pm$ 4000 \\
\end{tabular}
\end{center}
\caption{Correlation times in (integration time steps) for different
  wavelengths for $N=64$ particles.}
\label{tab:mean3}
\end{table}

In Tables~\ref{tab:mean2} and \ref{tab:mean3}, we show the
autocorrelation times for various modes in both Langevin and FAMD
simulations.  As seen from the tables, the FAMD dynamics is clearly more
efficient at decorrelating long wavelength modes.  A precise measure of
the gain over standard Langevin MD was not possible, because standard
Langevin MD has a very long correlation time.  We therefore do not know
exactly how much faster FAMD is.  Moreover, whether or not there is
simply a decrease in the amplitude of decorrelation time or an actual
decrease in its algebraic form is not known.  As with the precise
determination of $z$ for the $\phi^4$-model simulation, that will
require considerably more computational effort which we postpone to
future work.

Finally, we limited our investigation to a maximum of only 64 particles
to allow us to equilibrate the system at its critical point using
Langevin dynamics.  We expect, that gains over standard Langevin MD will
be ever more significant as the number of particles increases, both at
and away from the critical point.
 
\section{Discussion}
\label{sec:disc}

We have described a Fourier-based Langevin scheme, capable of
accelerating the dynamics of particulate systems with a Lagrangian
data representation.  We have demonstrated that there is great
potential in speeding up the dynamics of long-wavelength modes.  Two
issues related to the accelerated dynamics will be addressed in future
research.  (1) Dynamical faithfulness: how much of the actual dynamics
is unchanged by taking different time steps for different wavelength
modes?  (2) Does this method offer even more of a gain when it is
applied to molecular systems with nonconserved order parameters (for
example, dipolar systems or systems with structural phase
transitions)?  We believe that our preliminary computational
investigations Fourier methods have shown great promise, and we intend
to explore these questions in detail in future work.

\section*{Acknowledgements}

We thank Harvey Gould, Neils Gronbech-Jensen, and William Klein for very
useful discussions.  This work was supported in part by NSF grant
DMR-9633385, and under the auspices of the Department of Energy at Los
Alamos National Laboratory (LA-UR 00-118) under LDRD-DR 98605.  BMB was
partially supported by AFOSR Grant F49620-95-1-0285.

\end{document}